\documentclass[letterpaper,10pt]{article} 
\usepackage{osameet3} %% use version 3 for proper copyright statement

\usepackage{amsmath,amssymb}
\usepackage{xcolor}
\usepackage[colorlinks=true,bookmarks=false,citecolor=blue,urlcolor=blue]{hyperref} %pdflatex

\DeclareUnicodeCharacter{2212}{-}
\newcommand{\stefania}[1]{\textcolor{black}{#1}}

\begin{document}

\title{Silicon Carbide Metasurfaces for Controlling the Spontaneous Emission of Embedded Color Centers}
%\stefania {I think this title is misleading as we have no single photon emission, I would change in something more informative of what is done such us:
%Silicon Carbide Metasurfaces for Spontaneous Emission Rate Control}

\author{Mohammed Ashahar Ahamad and Faraz Ahmed Inam}
\affiliation{Department of Physics, Aligarh Muslim University, Aligarh, Uttar Pradesh 202002, India\\
}%
\email{faraz.inam.phy@amu.ac.in}
%\email{faraz.inam@gmail.com}

%\author{Faraz Ahmed Inam}
% \altaffiliation{Department of Physics, Aligarh Muslim University, Aligarh, U.P., India} %Lines break automatically or can be forced with \\

%\author{Nadeem Ahmed}%
% \email{khannadeemamu@gmail.com}
%\affiliation{%
% Department of Physics, Aligarh Muslim University, Aligarh, Uttar Pradesh 202002, India\\
%}%


\author{Stefania Castelletto}
\affiliation{School of Engineering, RMIT University, Melbourne, Victoria 3001, Australia\\ }
\email{Stefania.castelletto@rmit.edu.au}

%\affiliation{    School of Engineering, RMIT University, Melbourne, Victoria 3001, Australia\\
%}%
%\email{Stefania.castelletto@rmit.edu.au}

%\author{Faraz Ahmed Inam}
%\affiliation{Department of Physics, Aligarh Muslim University, Aligarh, Uttar Pradesh 202002, India}%

%\author{Faraz Ahmed Inam and Stefania Castelletto}
%\address{Department of Physics, Aligarh Muslim University, Aligarh, Uttar Pradesh, India, 20002}
%\address{School of Engineering, RMIT University, Melbourne, Victoria 3001, Australia}
%\email{faraz.inam@gmail.com}
%%Uncomment the following line to override copyright year from the default current year.
\copyrightyear{2022}

\begin{abstract}
Nanopillars  fabricated in diamond or silicon-carbide (SiC) have been used to enhance the light harvesting or absorption or to increase the collection efficiency of embedded single photon emission \stefania{in the visible or near infrared} for their detection using confocal microscopy. %Meta-lenses provide high focusing, large emission directionality and nearly unity collection efficiency for underneath color-centers. \stefania{ I think here we should not talk about metalens as the physical principle is totally different}
\stefania{While electric and magnetic dipolar resonances in SiC have been studied in the far-infrared, they have not been studied in the near infrared.  Here we show for the first time that electromagnetic Mie-scattering moments within SiC metasurfaces  can control the spontaneous emission process of \stefania{point defects in the near infrared}.} Using SiC nanopillars based metasurfaces, we theoretically demonstrate a control over the spontaneous emission rate of embedded color-centers by using the coherent superposition of the electric dipolar and magnetic quadrupolar electromagnetic Mie-scattering moments of the structure. More than an order of magnitude emission/decay rate enhancement is obtained with the maximum enhancement close to 30. We also demonstrate that the relative phase of the Mie-scattering moments helps in controlling the emission directionality. \stefania{SiC metasurfaces in the spectral range of color centres, from the visible to the near infrared, can be used to control the confinement and directionality of their spontaneous emission, increasing the opportunities to study light-matter interaction and to advance quantum photonic and quantum sensing device integration.}     

\end{abstract}

\section{Introduction}

Color-centers in silicon-carbide (SiC) are example of emitters that possess single photon emission\cite{Lohrmann2017}, optical spin read out and control, and have been amongst the most studied for optical coherence spin control and spin-photon interface\cite{awschalom2018quantum,son2020developing} due to their very long coherence time \cite{anderson2021five, Simin2017} and photo-stability. 
SiC is quite distinguished from the other material platforms as it possesses color centre with optical-spin properties combined with advanced material fabrication methods, metal-oxide-semiconductor functionalities \cite{liu2015silicon} and nonlinear second and third order optical properties. Due to its wide electronic bandgap which leads to broad optical transparency, photostable color centres emission\cite{castelletto2021silicon} which extend to the near infrared, \stefania{CMOS compatibility\cite{zhu2022hybrid} and availability of quantum-grade wafer-scale SiC on insulator, it has emerged as one of the most a promising material for integrated quantum photonic applications \cite{lukin2020integrated, castelletto2022silicon}.} 

In particular, SiC  can host a wide range of point defects/color centers including silicon vacancy $V_{\rm{Si}}$ (V1, V2, V3), divacancies $V_{\rm{Si}} V_{\rm{C}}$ and carbon antisite vacancy pair $C_{\rm{Si}} V_{\rm{C}}$\cite{castelletto2020silicon}. The V$_{\rm{Si}}$ in SiC is a promising single photon source (SPS) for spin-photon photon interface in the near infrared region around 917nm \cite{Radulaski2017,morioka2020spin}. At present the  main challenge in applying SiC for quantum networks is to significantly enhance the rate of single photon generation and collection from embedded color-centers. Photonics is mainly used to enhance the properties of these systems.

So far in SiC  bulk material, nanopillars have been fabricated to enhance the light harvesting or collection efficiency of embedded single photon emission for their detection using confocal microscopy \cite{Radulaski2017,Castelletto2019a}; while meta-lenses\cite{Schaeper2022} are used to \stefania{modify the phase front of the emitted light, achieving high focusing and large emission directionality of the color-centers emitting below these meta-lenses. 
Currently metasurfaces used to excite/enhance the magnetic and electric resonances in SiC have been investigated only in the far infrared \cite{Schuller2007} and in the context of surface phonon polaritons studies\cite{caldwell2013low}}. 
%For the last few decades, research in the field of quantum optics has advanced due to the availability of room temperature quantum systems in solids promising increased integration and scalability for development of quantum communication technology. One example of this technology consists in transferring the quantum information of electrons' spin to single photon \cite{morioka2020spin}, that can be used as spin read-out and/or flying qubits. 
Recently it has been shown that metamaterial/metasurface light matter interaction can be used to control, enhance and tune the quantum properties of bulk materials\cite{qiu2021quo,solntsev2021metasurfaces}. In particular all dielectric metasurfaces due to their zero absorption losses have emerged as the preferred platform compared to plasmonics in photo-luminescence enhancement \cite{LinHassanfirooziJiangLiaoLeeWu+2022+2701+2709}. When a dielectric structure is placed under electromagnetic excitation, various charge and current distributions are excited in it. These distributions results in multi-polar Mie-scattering resonances being excited in structures with dimensions of the order of the excitation wavelength \cite{Alaee2018}. A coherent superposition of these resonances leads to many interesting phenomena like, bound states in continuum (BIC) \cite{Liu2019}, tuning of the radiation directionality in the lateral or transverse directions \cite{Shamkhi2019a} and tuning of the local optical density of states (LDOS) to achieve emission rate enhancement for emitters embedded in the metasurfaces \cite{Khokhar2022}.

\stefania{Here for the first time we study the electric dipolar and magnetic quadrupolar resonances in the near infrared in SiC for controlling the spontaneous emission rate of the embedded color centers in the dielectric nanopillars forming Mie resonators.} 
%A control over the spontaneous emission of the embedded color-centers in these dielectric pillars can potentially accelerate their successful deployment in quantum technologies and sensing applications. 

In this study, using the coherent superposition of Mie-scattering resonances in SiC pillars based metasurface, we theoretically demonstrate that it is possible to control the spectral spontaneous emission process of the embedded color-centers.\stefania{ We first optimise the scattering efficiency of the SiC metasurface when excited by a plane wave and then by a dipole emitter. In the case the light source is a dipole, namely the V$_{\rm{Si}}$ embedded in the SiC metamaterial, we study the effects of the metasurface based on array of nanopillars Mie resonances on the LDOS and emission directionality. In particular, we study the effect of the periodicity of the nanopillars array to increase the emission rate and maintain high directionality compared to the case of a single pillar.}

\section{Theoretical background}
Scattering is the phenomenon of re-emission of radiation by a particle after undergoing interaction with radiation \cite{hulst1981light}. When a plane electromagnetic wave is incident on a particle, charge distribution and displacement currents  $J(\mathbf{r})  = -{i}{\omega}{\epsilon_{0}}({\epsilon_{r}}-1)\mathbf{E}(\mathbf{r})$ (here $\mathbf{E}{(\mathbf{r})}$ is the field at the position vector $\mathbf{r}$, $ {\omega} =  {2{\pi}r}$ is the angular frequency, ${\epsilon_{r}}$ and ${\epsilon_{0}}$ are the permittivity of the particle and surrounding medium) are excited within it. When the particle's dimensions are of the order of the excitation wavelength, the excited charge and current distributions leads to the development of multipolar Mie-scattering modes \cite{mie1976contributions, bezares2013mie}. The amplitude and phase of excitation of the electric and magnetic resonances or multi-polar Mie-scattering moments inside the scatterer are totally governed by its size, shape and surrounding electromagnetic environment \cite{bohren2008absorption}. These multi-polar Mie resonances in the visible spectral range have been demonstrated experimentally in the last decade using a silicon spherical nanopartciles and nanodiamonds\cite{evlyukhin2012demonstration,shilkin2017optical}.   

The total scattering efficiency (SE) \stefania{$C_{sca}^{total}$} is calculated by normalizing the total far field scattered power to the energy flux of the incident wave on the scatterer\cite{hinamoto2021menp}. The total SE $C_{sca}^{t}$ is the sum of partial SE from different multipoles:  $C_{sca}^{p}$, $C_{sca}^{m}$, $C_{sca}^{Q}$  and  $C_{sca}^{M}$ represents contributions from electric dipole, magnetic dipole, electric quadrupole and magnetic quadrupole respectively \cite{Alaee2015}.
\begin{eqnarray}
C_{sca}^{total} &=& C_{sca}^p+C_{sca}^m+C_{sca}^Q+C_{sca}^M  \\
C_{sca}^{total} &=& \frac{k^4}{6\pi \epsilon_0^2 |E_{inc}|^2} \left [ \sum \left ( |p_{\alpha}|^2 + \left |\frac{m_{\alpha}}{c} \right|^2 \right) + \frac{1}{120} \sum \left ( |kQ_{\alpha\beta}^{e}|^2 + \left |\frac{kQ_{\alpha\beta}^{m}}{c} \right|^2 \right)\right]
\label{Scattering_Eq}
\end{eqnarray}
where, $p_{\alpha}$ and $m_{\alpha}$ are the electric and magnetic dipole moments with $Q_{\alpha\beta}^{e}$ and $Q_{\alpha\beta}^{m}$ being the corresponding quadrupole moments. $|E_{inc}|$ is the amplitude of the incident electric field, $k$ is the wave-vector and $c$ is the speed of light. They are mathematically expressed as \cite{Alaee2015}:
\begin{eqnarray}
ED:  p_{\alpha}  &=& -\frac{1}{i\omega}\Bigl\{\int{d}^{3}\mathbf{r}{J}_{\alpha}^{\omega}j_{0}(kr)+\frac{k^2}{2}\int{d}^{3}\mathbf{r}\left[3(\mathbf{r}.\mathbf{J}_{\omega}){r}_{\alpha}-{r}^2{J}_{\alpha}^{\omega}\right]\frac{j_{2}(kr)}{(kr)^2}\Bigr\} \\
MD: m_{\alpha} &=& \frac{3}{2}\int{d}^{3}\mathbf{r}(\mathbf{r}\times \mathbf{J}_{\omega})_{\alpha}\frac{j_{1}(kr)}{kr} \\
EQ: Q_{\alpha\beta}^{e} &=& -\frac{3}{i\omega}\Bigl\{\int{d}^{3}\mathbf{r}[3(r_{\beta}{J}_{\alpha}^{\omega}+r_{\alpha}{J}_{\beta}^{\omega})-2(\mathbf{r}.\mathbf{J}_{\omega}){\delta}_{\alpha\beta}]\frac{j_{1}(kr)}{(kr)} \nonumber\\ &+&   2k^2\int{d}^{3}[5r_{\alpha}r_{\beta}(\mathbf{r}.\mathbf{J}_{\omega})-(r_{\alpha}J_{\beta}+r_{\beta}J_{\alpha})r^2 - r^2(\mathbf{r}.\mathbf{J}_{\omega}){\delta}_{\alpha\beta}]\frac{j_{3}(kr)}{(kr)^3}\Bigr\}  \\
MQ: Q_{\alpha\beta}^{m} &=& 15\int{d}^{3}\mathbf{r}\Bigl\{r_{\alpha}(\mathbf{r}\times \mathbf{J}_{\omega})_{\beta} + r_{\beta}(\mathbf{r}\times \mathbf{J}_{\omega})_{\alpha}\Bigr\}\frac{j_{2}(kr)}{{(kr)^2 }}
\end{eqnarray}

The Mie-resonances control the electromagnetic field amplitudes within the scatterer and therefore contribute in tuning the local electromagnetic density of states (LDOS). The LDOS due to the local electromagnetic environment around a point dipole emitter is defined as \cite{inam2011modification}
\begin{equation}
    \rho(\omega ,r)=\sum_{k,\sigma} |\hat{d} \cdot \mathbf{E}_{k,\sigma}(r)|^2 \delta (\omega-\omega_{k,\sigma}).
    \label{LDOS_Eq}
\end{equation}
Here, $\hat{d}$ is the unit vector specifying the direction of the transition dipole moment with $\omega$ being the transition frequency. The summation is over all wavevectors (k) and polarizations ($\sigma$). \textbf{E} is the total electric field at the source position \stefania{resulting from the superposition of the fields directly radiated by the dipole emitter embedded inside the scatterer with the fields reflected and scattered back from the surroundings.} The LDOS govern the complete radiation process of a dipole emitter. \stefania{Hence the Mie-scattering modes} play a vital role in tuning the spontaneous emission process of the emitter by controlling the scattered electric field at the source point.
\begin{figure}[t]
    \centering
    \includegraphics[width=5.25in]{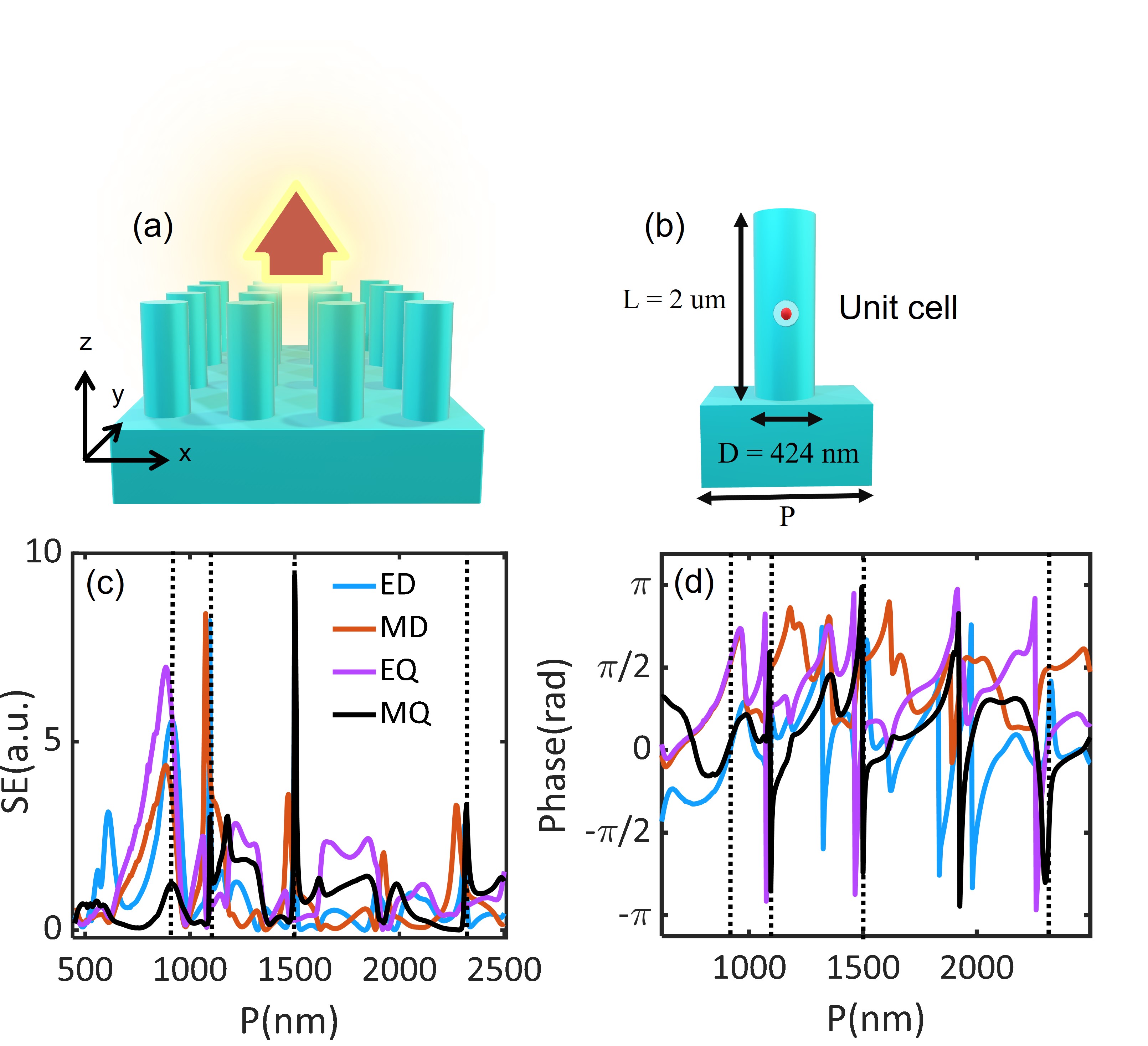}
    \caption{(a) Schematic of the metasurface with a 2D periodic lattice of SiC pillars under plane-wave excitation at 917 nm with wave-vector along the +z-direction. (b) Schematic of the unit cell. Each pillar has a length L = 2 $\mu$m and diameter D = 424 nm with a dipole emitter located at the center of each pillar. (c) The SE of the individual multipolar Mie-scattering moments as a function of the lattice periodicity, $P$, under plane-wave excitation at 917 nm. $P$ is varied from 450 nm to 2500 nm. (d) The corresponding phase of the individual multipolar Mie-scattering moments as a function of the lattice periodicity, $P$. The dotted black lines corresponds to overlapping ED and MQ resonances with $P =$ 915 nm, 1095 nm, 1500 nm and 2315 nm.}
    \label{schematic_metasurface}
\end{figure}

The balancing of the electric and magnetic Mie-scattering moments leads to the directionality of the \stefania{scattered radiation pattern} \cite{Alaee2015}. The radiation pattern is controlled by the relative phase of the balanced electric and magnetic multi-polar moments \cite{shamkhi2019transverse}. When the electric and magnetic dipolar moments are balanced and in phase, $|ED| = |MD|$, $arg(ED) = arg(MD)$, this leads to a completely forward radiation directionality, known as the Kerker condition \cite{Alaee2015}. When these dipolar moments are out-of-phase, $|ED| = |MD|$, $arg(ED) = arg(MD) + \pi$, it results in a completely backward directionality, known as the anti-Kerker condition \cite{shamkhi2019transverse}. When the superposition of dipolar as well as the quadrapolar moments are balanced and are in phase, $|ED + MD| = |EQ + MQ|$ with $arg(ED + MD) = arg(ED + MD)$, the radiation pattern is highly directional along the forward direction, known as the generalised Kerker condition \cite{shamkhi2019transverse}. However, when these superpositions are out-of-phase, this leads to a complete transverse scattering \cite{shamkhi2019transverse}. The phase of the Mie-scattering moments therefore controls the far-field scattering radiation pattern of the scatterer. Under a point dipole emitter excitation of the structure, the far-field scattering pattern of the structures also influences the radiation pattern of the dipole emitter placed in the vicinity of the scatterer \cite{Feifei2022PRL}.

\stefania{In the following we investigate these effects in SiC nanopillars array under plane wave-excitation and under a single dipole excitation simulating the $V_{\rm Si}$.}

%In this study, we designed a metasurface with a periodic 2D lattice of SiC pillars. MieThe scattering efficiency is further enhanced due to coherent interaction of pillar-to-pillar modes. The superposition of these multipolar Mie-scattering modes results in resonance maxima. This increase the electric field intensities in the pillars and this in-turn leads to the enhancement of the local density of optical state (LDOS) within each pillar  \cite{khokhar2022kerker}. The LDOS controls the spontaneous emission rate or the modified Purcell factor, $F_p$ ($F_p = \gamma/\gamma_{\infty}$, $\gamma$ being the decay rate of the emitter in the pillar with $\gamma_\infty$ being the decay rate in bulk SiC) for emitter embedded within the pillar. 

\section{Results and discussion}
\begin{figure}
\centering
\includegraphics[width=\textwidth]{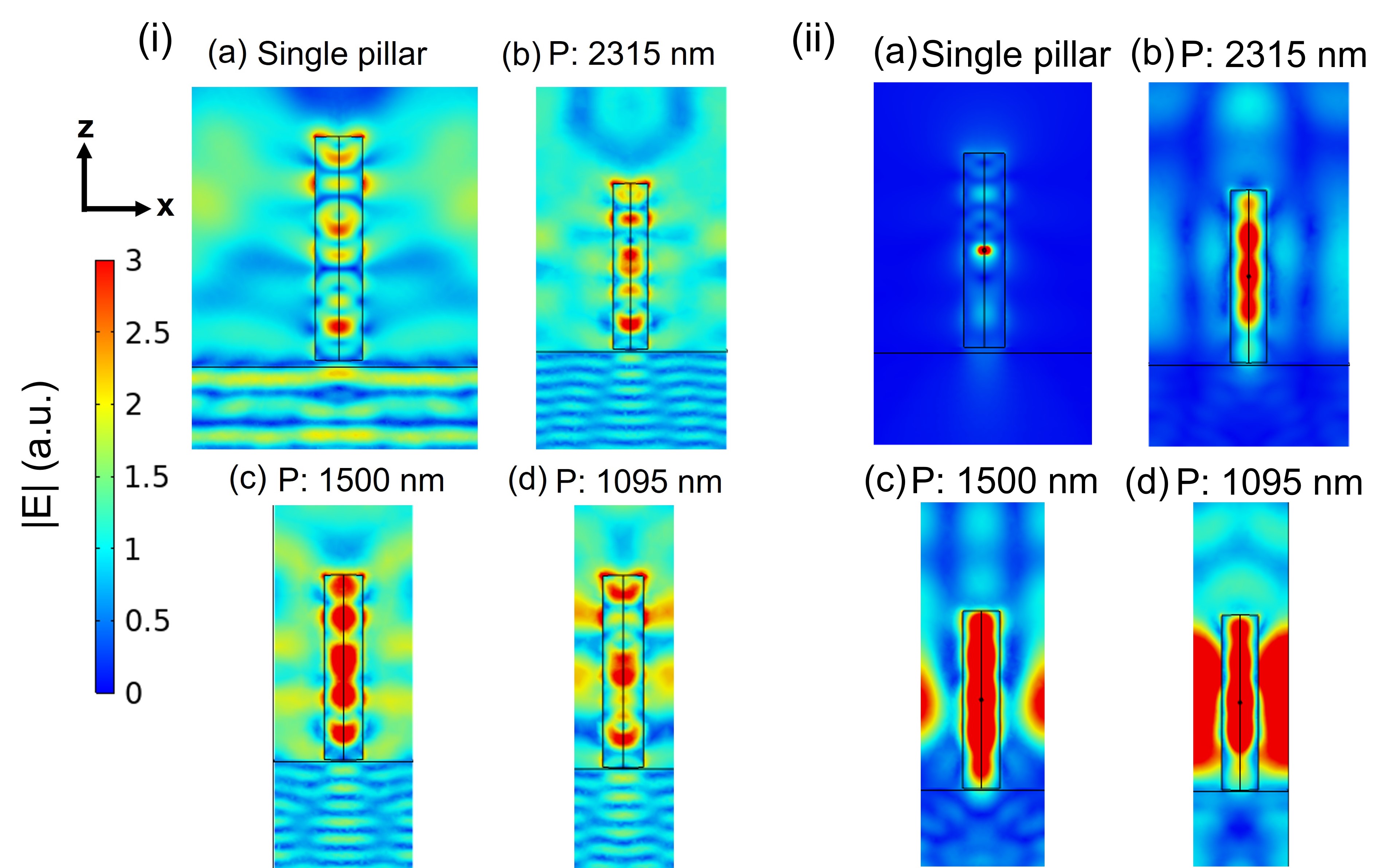}
\caption{The 2D electric field norm within (a) a single SiC pillar and the SiC pillar metasurface with $P=$ (b) 2315 nm, (c) 1500 nm and (d) 1095 nm under (i) plane wave excitation with wave-vector along the +z-direction and electric field polarized along the +x-direction and (ii) dipole excitation with dipole emitter placed at the center of the SiC pillar with orientation along the x-direction.}
\label{Efield_profile}
\end{figure}
\begin{figure}[t]
\centering
\includegraphics[width=\textwidth]{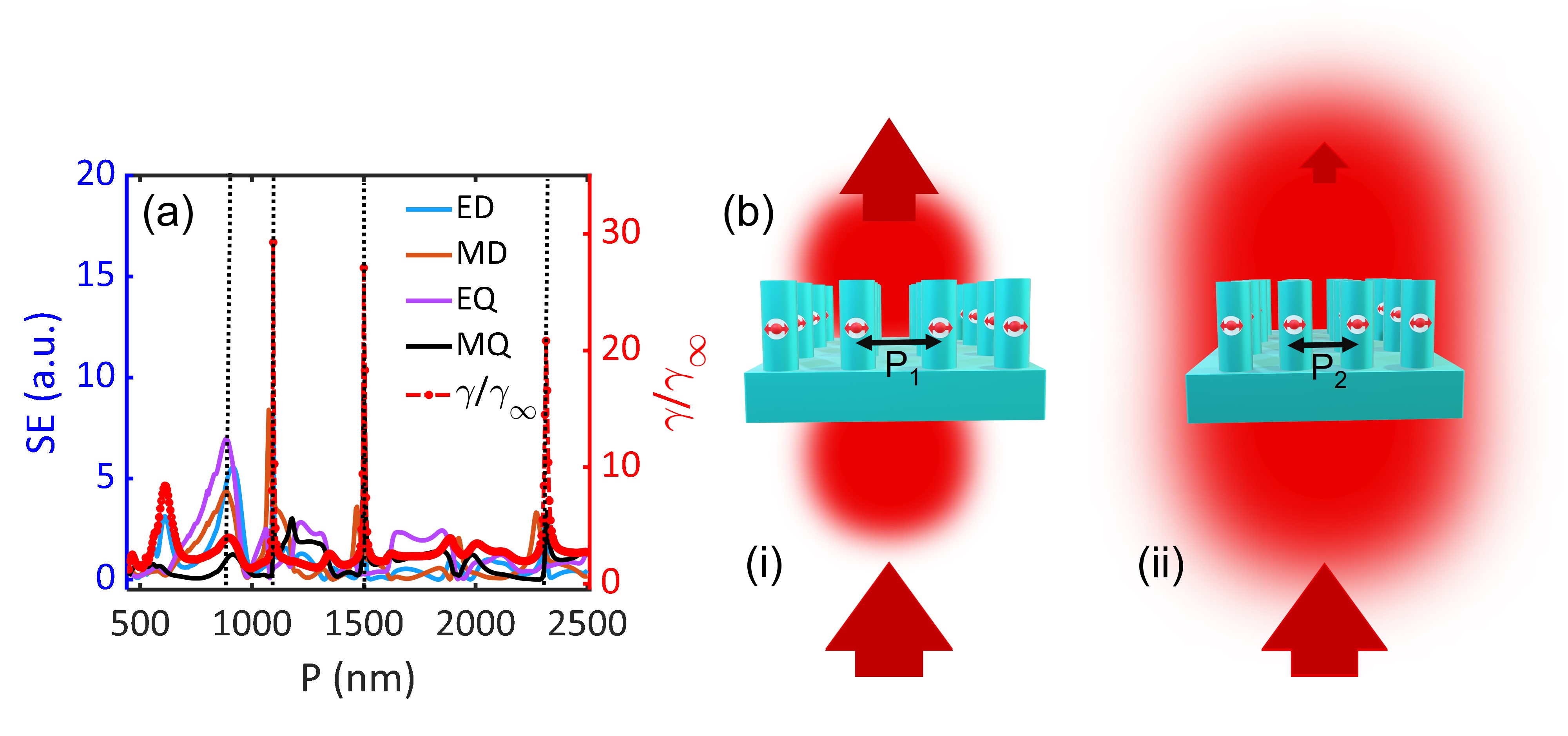}
\caption{The SE of the individual excited multipolar Mie-scattering moments and the emitter's ($V_{\rm{Si}}$ color-center) relative decay rate in the SiC pillar metasurface as a function of the lattice periodicity, $P$. The dipole emitter is placed at the center of the SiC pillars with dipole orientation along the horizontal plane. The $\gamma_{\infty}$ is the emitter's decay rate in the bulk SiC. (b) The schematic representation of the tuning of the embedded color-center's emission with the lattice periodicity, P set to (i) off-resonant $P_1$ and (ii) resonant $P_2$ values.}
\label{SE_Relative_rate_with_P}
\end{figure}

\subsection{Scattering efficiency and decay-rate enhancement}
\stefania{ We have computationally optimised the SiC} metasurface, shown in Fig.~\ref{schematic_metasurface}(a) with unit cell in Fig.~\ref{schematic_metasurface}(b), \stefania{to achieve the generalised Kerker's condition in SiC for the specific color centre of interest}. The metasurface consists of a periodic 2D lattice of SiC pillars, each of length, $L = 2 \mu$m. The electrodynamics calculations are performed using the commercial Comsol Multiphysics RF module. The details of the calculations are presented in the Methods sections. The metasurface is excited by a plane wave with wavelength, $\lambda_{exc}$, travelling along the +z-direction (arrow symbol in Fig.~\ref{schematic_metasurface}(a)) with the electric field polarized along the +x-direction. Under the influence of the plane electromagnetic wave, Mie scattering moments are excited within the SiC pillars. We first optimised the diameter, $D$, of the SiC pillars for the maxima in the SE at $\lambda_{exc} = 917$ nm corresponding to the zero phonon line (ZPL) of the silicon vacancy, $V_{\rm{Si}}$ in SiC \cite{Fuchs2015,Widmann2015}. The optimised $D$ value was found to be around $424$ nm. We then study the coherent superposition of the Mie-scattering modes of the individual SiC pillars by varying the lattice periodicity, $P$. For $P \gg \lambda_{exc}$, the structure is expected to behave as a single isolated pillar. With decreasing $P$, the interactions between the Mie-scattering modes of the individual pillars will increase. When these modes will be in phase, their coherent superposition will lead to a maxima for the total SE of the 2D SiC pillar lattice. Fig.~\ref{schematic_metasurface}(c) and (d) show the amplitude and the phase of the individual Mie-scattering moments of the SiC pillar metasurface as a function of $P$. Sharp resonance peaks are observed in the amplitude of the individual Mie-scattering moments (Fig.~\ref{schematic_metasurface}(c)). At these sharp resonances, a sharp jump in the phase of the corresponding Mie-scattering moment is observed (Fig.~\ref{schematic_metasurface}(d)). We will now focus on the local maxima arising due to the ED and the MQ moments (under dipole excitation of the structure only these two resonances were excited and were observed to have an influence on the dipole emitter's decay rates). These local maxima are observed for $P = $915 nm, 1095 nm, 1500 nm and 2315 nm (black dotted lines in Fig.~\ref{schematic_metasurface}(c) and (d)). At $P = $915 nm, the ED resonance peak is much greater than the MQ resonance with the phase of these two resonances being equal. For $P = $1095 nm, 1500 nm and 2315 nm, the ED and MQ moments are nearly balanced and a sharp jump is also observed in their phase. We will now  \stefania{examine the balanced superposition of these two moments at $P = $1095 nm, 1500 nm and 2315 nm. }    

Figure~\ref{Efield_profile}(i) shows the normalised electric field distribution within a (a) single SiC pillar and the SiC pillar metasurface with $P =$ (b) 2315 nm, (c) 1500 nm and (d) 1095 nm under plane wave excitation with wave-vector along the +z-direction and electric field polarized along the +x-direction. Strong confinement of the electric field within the SiC cylinder is observed at these $P$ values corresponding to the balanced superposition of the ED and MQ resonances, with the maximum field confinement observed for $P =$ 1500 nm (the SE was also observed to be maximum at this P value). The field confinement will in-turn lead to LDOS enhancement within the SiC pillar (Eq~\ref{LDOS_Eq}). For a dipole emitter placed at the field maxima points, the LDOS enhancement will lead to its decay rate enhancement. Figure Fig.~\ref{Efield_profile}(ii) shows the normalised electric field distribution within (a) a single SiC pillar and the SiC pillar metasurface with $P =$ (b) 2315 nm, (c) 1500 nm and (d) 1095 nm under dipole excitation with dipole emitter placed at the center of the SiC pillars with orientation along the x-direction. Large field confinement/enhancement which will lead to large LDOS enhancement can be observed here.  

We now study the influence of the above LDOS enhancement on the spontaneous emission rates of a dipole emitter, the  $V_{\rm{Si}}$ color-center embedded at the center of each SiC pillar. Figure~\ref{SE_Relative_rate_with_P}(a) shows the emitter's ($V_{\rm{Si}}$ color-center) relative decay rate together with the SE of the individual Mie-scattering moments in the SiC pillar metasurface as a function of the lattice periodicity, $P$. The  decay rates of the  $V_{\rm{Si}}$ emitter in the SiC pillar metasurface, $\gamma$ are scaled relative to its decay rates in a bulk SiC crystal, $\gamma_{\infty}$. The influence of the LDOS enhancement arising from the electric field confinement (Fig.~\ref{Efield_profile}) in tuning the emitter's decay rate can be clearly observed here. Also, it can be observed that the relative decay rates ($\frac{\gamma}{\gamma_{\infty}}$) (dash-dotted red curve) only tunes with the local maxima which are dominated by ED and MQ resonances. These resonances corresponds to $P =$ 1095 nm, 1500 nm and 2315 nm, respectively. A schematic representation of the embedded dipole emitter's radiation tuning with the SiC pillar lattice periodicity, $P$ at an off-resonant (i) and resonant (ii) value is presented in Fig.~\ref{SE_Relative_rate_with_P}.
\begin{figure}[t]
    \centering
    \includegraphics[width=0.8\textwidth]{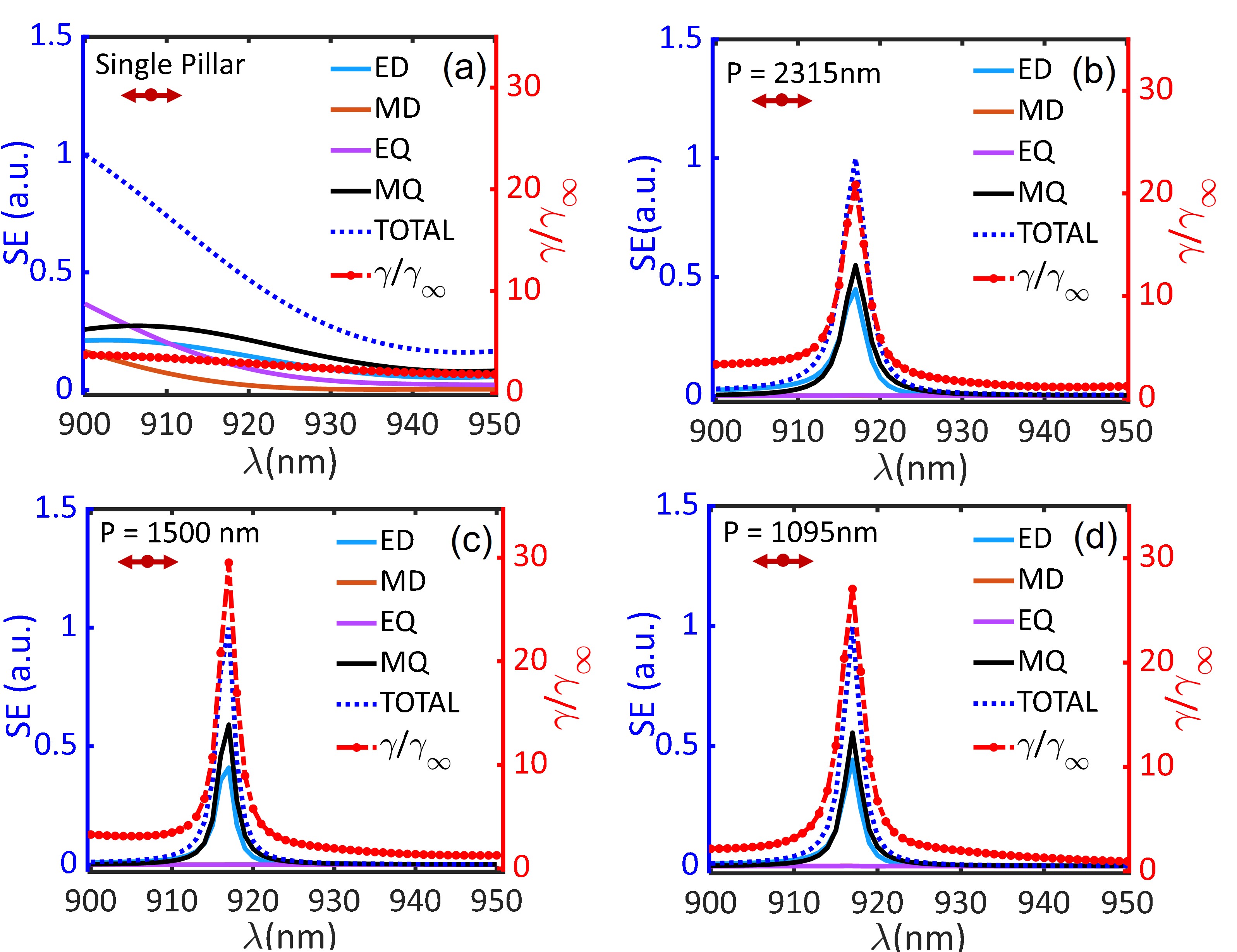}
    \caption{The spectral response of the SE with the individual excited multipolar Mie-scattering moments under horizontal dipole excitation and the emitter's ($V_{\rm{Si}}$ color-center with dipole orientation along the horizontal plane) relative decay rate in a (a) single SiC pillar; SiC pillar metastuface with $P =$ (b) 2315nm, (c) 1500 nm and (d) 1095 nm, respectively.}
    \label{SE_rate_spectrum}
\end{figure}

In Fig.~\ref{SE_rate_spectrum}, we study the SE spectral response due to all the excited Mie-scattering moments and \stefania{the effect on} the relative decay rates of a horizontally oriented (along x-direction) dipole source for the above resonant periodicity values ($P =$ 1095 nm, 1500 nm and 2315 nm) of the metasurface. Here, the Mie-scattering moments of the metasurface are excited by the dipole source itself. For an isolated SiC pillar (Fig.~\ref{SE_rate_spectrum}(a)), all the studied Mie-scattering moments are observed to be weakly excited with no clear resonances. The relative decay rate (dash-dotted red curve) is observed to be around 2.7 at 917 nm. However, for all resonant $P$ values ($P =$ 1095 nm, 1500 nm and 2315 nm), significant contributions are observed only from the ED and MQ moments. Their superposition is controlling the behaviour of the SE and the relative decay rate. The maximum relative decay rate enhancement is  close to 30 at 917 nm for $P =$ 1500 nm and 1095 nm. For $P =$ 2315 nm the enhancement is about 20. Therefore, it can be concluded that the coherent superposition of the ED and MQ Mie-scattering moments of the individual pillars are enhancing the decay rates of an embedded dipole emitter by more than an order of magnitude. 

We now study the role of the phase of the excited Mie-scattering moments (ED and MQ) of the SiC pillar metasurface on the far field radiation pattern of an embedded dipole emitter. 
\begin{figure}[t]
\centering
\includegraphics[width=0.8\textwidth]{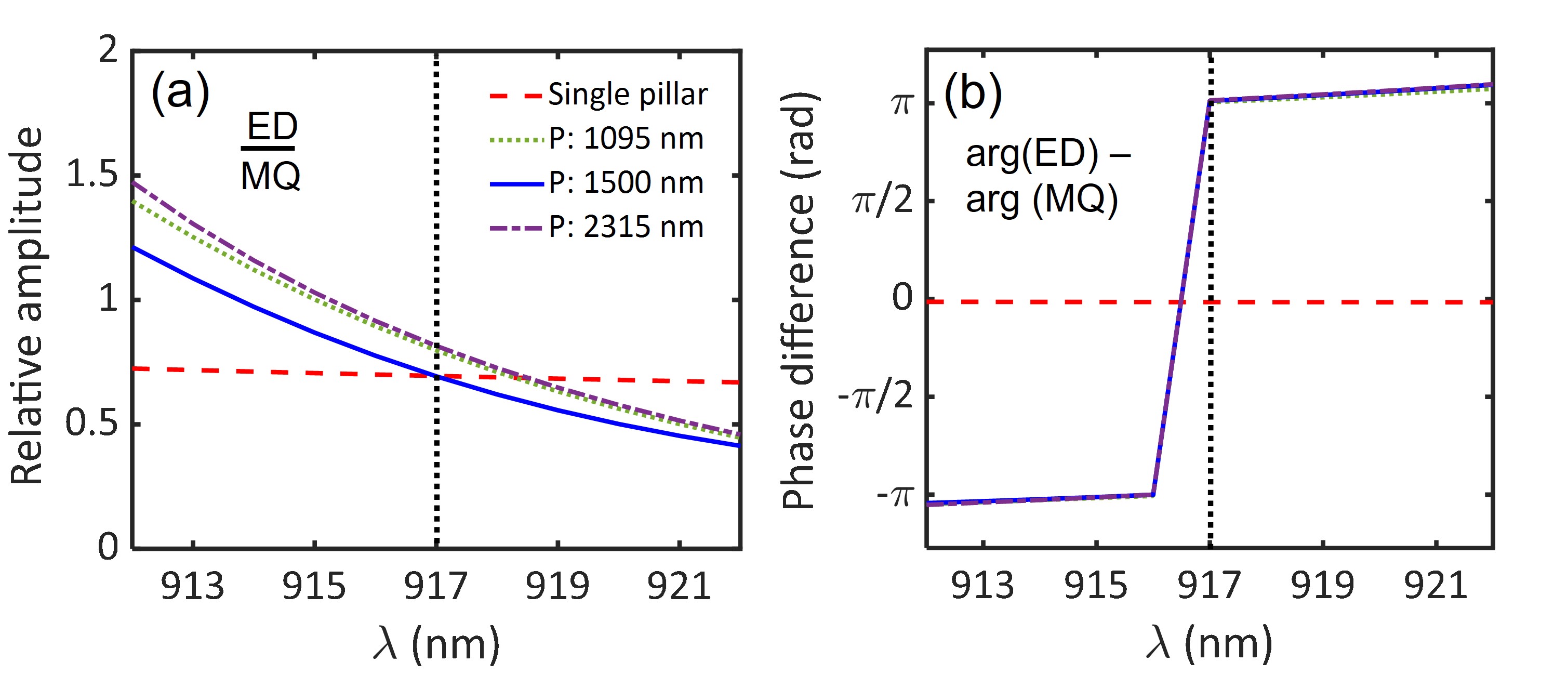}
\caption{The spectral response of the (a) relative amplitudes and (b) phase difference of the electric dipolar (ED) and magnetic quadrapular (MQ) Mie-scattering moments excited under horizontal dipole excitation.}
\label{rel_amp_phase}
\end{figure}
\begin{figure}[t]
\centering

\includegraphics[width=0.8\textwidth]{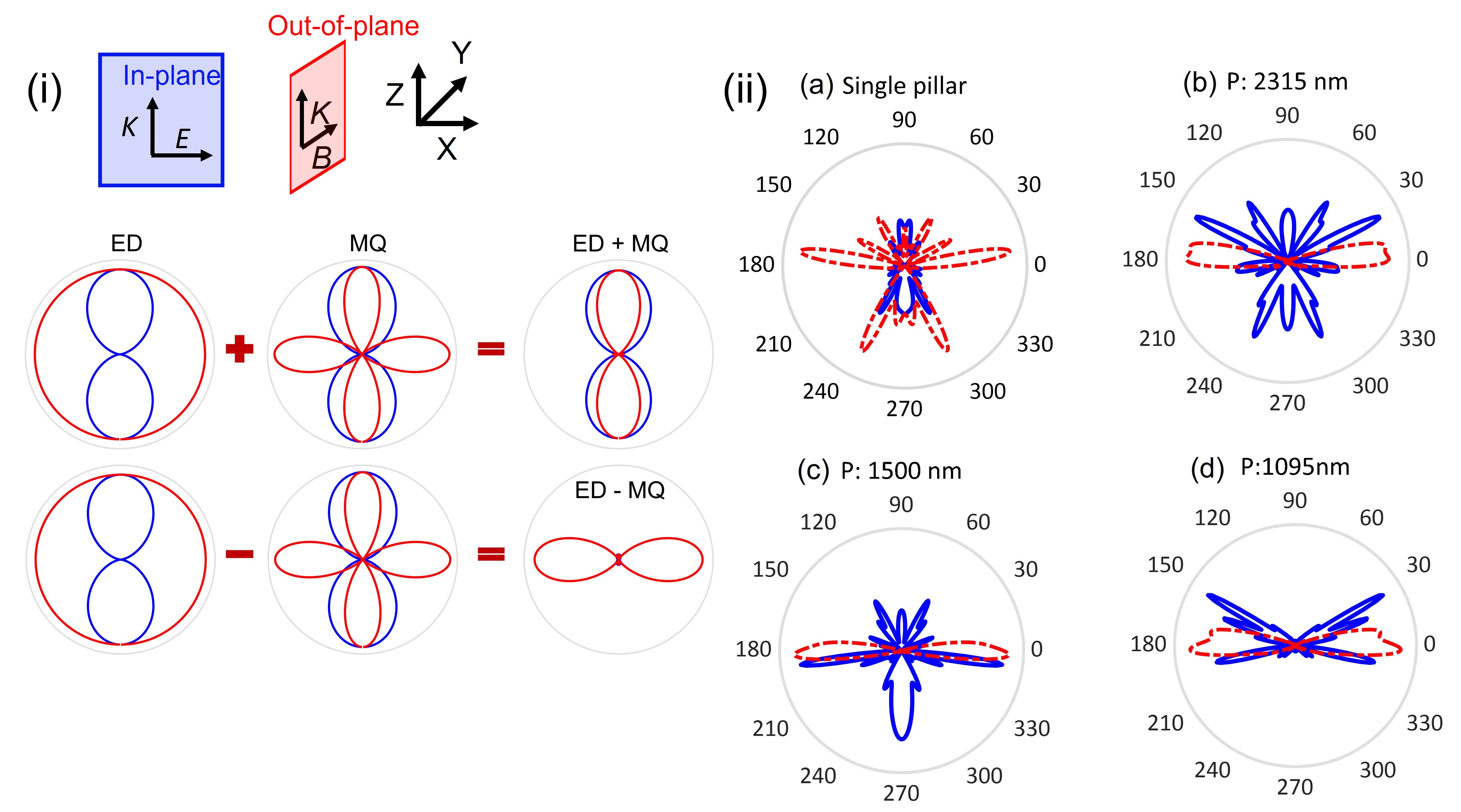}
\caption{(i) In-plane (X-Z plane, blue curves) and out-of-plane (Y-Z plane, red curves) far-field scattering pattern corresponding to the in-phase (\stefania{ED+MD}) and out-of-phase \stefania{(ED-MQ)} superposition of the electric dipolar (ED) and magnetic quadrupolar (MQ) Mie-scattering moments. (ii) Farfield radiation patterns in-plane (X-Z plane, blue curves) and out-of-plane (Y-Z plane, red curves) of a dipole emitter (orientation along the X-direction, same as that of electric field in (i)) placed at the center of (a) single pillar; SiC pillar metastuface with $P =$ (b) 2315nm, (c) 1500 nm and (d) 1095 nm, respectively.}
\label{ED_MQ_superpositionandFarfield}
\end{figure}

\subsection{Phase analysis and Radiation pattern}

Figure~\ref{rel_amp_phase} shows a \stefania{narrow range of values for both the relative amplitudes and phase} of the ED and MQ moments at the resonant $P$ values. At the $V_{\rm{Si}}$ color-center's peak emission wavelength of 917 nm, the MQ moment \stefania{appears to be slightly larger than the ED moment}. For $P =$ 1500 nm, the $\frac{ED}{MQ} = 0.7$ and for $P =$ 1095 nm and 2315 nm, the $\frac{ED}{MQ} = 0.81$. The corresponding phase difference between ED and MQ moments is $\pi$ at 917 nm for all resonant $P$ values.  

In Fig.~\ref{ED_MQ_superpositionandFarfield}(i) we \stefania{show} the in-phase and out-of-phase superposition of the balanced ED and MQ moments. The influence of this superposition on both the in-plane (X-Z plane, blue curve) and out-of-plane (Y-Z plane, red curve) far-field scattering patterns is shown here. The in-phase ($ED + MQ$) superposition \stefania{results} in longitudinal (along top and bottom directions) scattering and the out-of-phase ($ED - MQ$) superposition results in transverse scattering along the out-of-plane direction (red curve). 

We now study the influence of the ED and MQ moments superpositions on the embedded dipole emitter radiation pattern \stefania{for a single nanopillar and SiC metasurface of array of interacting nanopillars}. The emitter's dipole orientation is along the X-direction. Fig.~\ref{ED_MQ_superpositionandFarfield}(ii) shows the far-field emission patterns of the embedded $V_{\rm{Si}}$ color-center at 917 nm for different resonant $P$ values. In a single pillar, the majority of the emission is observed to be directed towards the bottom surface and lies mainly along the longitudinal direction (bottom and top). However, in the SiC metasurface at the resonant $P$ values corresponding to ED and MQ moments, the out-of-phase superposition of the ED and MQ moments (the phase difference between these moments was observed to be close to $\pi$ at 917 nm in Fig.~\ref{rel_amp_phase}) \stefania{directs} the embedded emitter's radiation pattern along the transverse direction, especially in the out-of-plane (Y-Z plane, red curve). 

\section {Conclusion} 
We studied for the first time the coherent superposition of the \stefania{eletric and magnetic dipolar and quadrupolar} Mie-scattering moments of SiC \stefania{metamaterial nanopillars array in the near infrared emission (917 nm)}. We first determined the design of the metasurface periodicity to induce sharp resonances in the amplitude and phase of the Mie-scattering moments. Strong electric field confinement was observed within the SiC pillar when the periodicity of the lattice matched with the resonance of the ED and MQ modes. The field confinement leads to large LDOS and subsequently decay rate enhancement for a color-center dipole embedded at the center of the SiC pillar. Under a point dipole emitter excitation within SiC, it was determined that only the ED and MQ moments are contributing to the electromagnetic scattering in the SiC nanopillars metasurface. Both these moments were observed to be nicely coupled and were showing collective resonance at the optimised wavelength (917 nm). The coherent superposition of these two moments controls the complete spontaneous emission process of the embedded color center. At the collective resonance point of these two moments ($\lambda = 917$ nm), we determined more than an order of magnitude decay rate enhancement with the maximum enhancement reaching 30. Such an enhancement has never been reported in dielectric neither in metal-dielectric individual nanopillars\cite{InamCastnano22}, thus paving the way for the use SiC metasurfaces to to enhance and control light extraction from quantum emitters, to study light matter interaction effects in integrated quantum photonics and for applications in quantum sensing. We also observed that by designing specific resonant structures, the coherent superposition of the ED and MQ moments can be used to better control the radiation/emission pattern and hence the emission directionality of the embedded dipole emitter compared to a single nanopillar. Specifically the embedded emitter's radiation pattern can be more confined along the transverse direction, especially in the out-of-plane (Y-Z plane, red curve), \stefania{ thus facilitating the planar emission propagation. Such result is relevant for applications of SiC metasurfaces for planar integrated photonics. Our study can prompt further studies on SiC quantum states of light based on multi-emitters-resonators coupling enabled by metamaterials such as superradiance\cite{Mello2022}.}% and studies on SiC epsilon near zero properties\cite{Kim:16} in the spectral region of SiC color centres due to their relevance for on chip quantum networks\cite{vertchenko2019epsilon}. }

\section{Materials and Methods}
All the electrodynamics calculations have been performed using the commercial COMSOL Multiphysics Radio Frequency (RF) module. The periodic boundary conditions are applied to all the horizontal planes defining the 2D lattice of the SiC substrate to build an array of dielectric pillars metasurface. The Scattering boundary conditions are applied at the top and bottom boundaries of the computational domain. The optical constant for the SiC has been extracted from the experimentally reported values by Singh et.al. \cite{singh1971nonlinear}. During the entire calculation the minimum meshing size was 1 nm with the maximum being $\lambda/7$. 

\subsection{Scattering efficiency calculation}
The scattering cross section is defined as the amount of power scattered by the scatterer to the amount of power per unit area carried by the incident wave. The SE is obtained just by dividing the scattering cross-section by the geometrical cross-section. Mathematically it is expressed as $SE = \sigma_s/G$\cite{frezza2017tutorial}. Here $\sigma_s$ is the scattering cross section and $G$ is geometrical cross section.

The SE is calculated semi-analytically using electric field values at each mesh point in the computational grid under plane-wave excitation using Comsol Multi-physics module. Using these field values and permittivity profile at each mesh points, current density is calculated as: $J_{\omega}(r) = i\omega\epsilon_0(\epsilon_r − 1)E_{\omega}(r)$. Here $\epsilon_0$  and $\epsilon_r$ are permittivity of free space and SiC medium, respectively. The computationally obtained values of $E(r)$, $J_{\omega}(r)$ and $\epsilon(r)$ are used to calculate individual multipolar Mie-scattering moments, $p_{\alpha}$, $m_{\alpha}$, $Q^e_{\alpha\beta}$ and $Q^m_{\alpha\beta}$ described in Eq.~\ref{Scattering_Eq}. The integration referred in Eq.~\ref{Scattering_Eq} is carried on the domain of the SiC pillar.

\subsection{Relative decay rate calculations}
In these calculations, the SiV center is treated as a classical radiating point dipole source. In the computational domain, it is modelled as a point current source driven at the emission frequency, $\nu = \frac{c}{\lambda}$ \cite{Xu1999}. Scattering/perfectly matched layer (PML) boundary conditions are applied on the exterior boundaries of the computational domain. The total power radiated by the dipole is integrated over a closed surface enclosing the current source. The relative decay-rate is calculated as $\Gamma_{rel}=\gamma/\gamma_{\infty} = P/P_{\infty}$ \cite{Xu1999}, where $P_{\infty}$ is the power corresponding to the point dipole's emission in the bulk SiC. The permittivity of SiC is taken from \cite{Shaffer1971}.\\

\textbf{Acknowledgement}
The authors would like to acknowledge the financial support from the Department of Science and Technology (DST), India (CRG/2021/001167). The authors thank Dr Nadeem Ahmed for his help regarding the plotting of the figures and the formatting of the manuscript. MA and FAI also thank Dr Ahmed Mekawy for his help regarding multipole decomposition of the Mie-scattering moments.

%\bibliography{My_Collection.bib}
\bibliographystyle{osajnl}
\bibliography{My_Collection}

\begin{thebibliography}{10}
\newcommand{\enquote}[1]{``#1''}

\bibitem{Lohrmann2017}
A.~Lohrmann, B.~C. Johnson, J.~C. McCallum, and S.~Castelletto, \enquote{{A
  review on single photon sources in silicon carbide},}
  {\protect\JournalTitle{Reports on Progress in Physics}} \textbf{80}, 034502
  (2017).

\bibitem{awschalom2018quantum}
D.~D. Awschalom, R.~Hanson, J.~Wrachtrup, and B.~B. Zhou, \enquote{Quantum
  technologies with optically interfaced solid-state spins,}
  {\protect\JournalTitle{Nature Photonics}} \textbf{12}, 516--527 (2018).

\bibitem{son2020developing}
N.~T. Son, C.~P. Anderson, A.~Bourassa, K.~C. Miao, C.~Babin, M.~Widmann,
  M.~Niethammer, J.~Ul~Hassan, N.~Morioka, I.~G. Ivanov \emph{et~al.},
  \enquote{Developing silicon carbide for quantum spintronics,}
  {\protect\JournalTitle{Applied Physics Letters}} \textbf{116}, 190501 (2020).

\bibitem{anderson2021five}
C.~P. Anderson, E.~O. Glen, C.~Zeledon, A.~Bourassa, Y.~Jin, Y.~Zhu,
  C.~Vorwerk, A.~L. Crook, H.~Abe, J.~Ul-Hassan, T.~Ohshima, N.~T. Son,
  G.~Galli, and D.~D. Awschalom, \enquote{Five-second coherence of a single
  spin with single-shot readout in silicon carbide,}
  {\protect\JournalTitle{Sci. Adv.}} \textbf{8}, eabm5912 (2022).

\bibitem{Simin2017}
D.~Simin, H.~Kraus, A.~Sperlich, T.~Ohshima, G.~V. Astakhov, and V.~Dyakonov,
  \enquote{{Locking of electron spin coherence above 20 ms in natural silicon
  carbide},} {\protect\JournalTitle{Phys. Rev. B Condens. Matter.}}
  \textbf{95}, 161201 (2017).

\bibitem{liu2015silicon}
G.~Liu, B.~R. Tuttle, and S.~Dhar, \enquote{Silicon carbide: A unique platform
  for metal-oxide-semiconductor physics,} {\protect\JournalTitle{Applied
  Physics Reviews}} \textbf{2}, 021307 (2015).

\bibitem{castelletto2021silicon}
S.~Castelletto, \enquote{Silicon carbide single-photon sources: challenges and
  prospects,} {\protect\JournalTitle{Materials for Quantum Technology}}
  \textbf{1}, 023001 (2021).

\bibitem{zhu2022hybrid}
Y.~Zhu, W.~Wei, A.~Yi, T.~Jin, C.~Shen, X.~Wang, L.~Zhou, C.~Wang, W.~Ou,
  S.~Song \emph{et~al.}, \enquote{Hybrid integration of deterministic quantum
  dots-based single-photon sources with cmos-compatible silicon carbide
  photonics,} {\protect\JournalTitle{arXiv preprint arXiv:2203.12202}}  (2022).

\bibitem{lukin2020integrated}
D.~M. Lukin, M.~A. Guidry, and J.~Vu{\v{c}}kovi{\'c}, \enquote{Integrated
  quantum photonics with silicon carbide: challenges and prospects,}
  {\protect\JournalTitle{PRX Quantum}} \textbf{1}, 020102 (2020).

\bibitem{castelletto2022silicon}
S.~Castelletto, A.~Peruzzo, C.~Bonato, B.~C. Johnson, M.~Radulaski, H.~Ou,
  F.~Kaiser, and J.~Wrachtrup, \enquote{Silicon carbide photonics bridging
  quantum technology,} {\protect\JournalTitle{ACS Photonics}} \textbf{9},
  1434--1457 (2022).

\bibitem{castelletto2020silicon}
S.~Castelletto and A.~Boretti, \enquote{Silicon carbide color centers for
  quantum applications,} {\protect\JournalTitle{Journal of Physics: Photonics}}
  \textbf{2}, 022001 (2020).

\bibitem{Radulaski2017}
M.~Radulaski, M.~Widmann, M.~Niethammer, J.~L. Zhang, S.-Y. Lee, T.~Rendler,
  K.~G. Lagoudakis, N.~T. Son, E.~Janz{\'{e}}n, T.~Ohshima, J.~Wrachtrup, and
  J.~Vu{\v{c}}kovi{\'{c}}, \enquote{{Scalable Quantum Photonics with Single
  Color Centers in Silicon Carbide},} {\protect\JournalTitle{Nano Letters}}
  \textbf{17}, 1782--1786 (2017).

\bibitem{morioka2020spin}
N.~Morioka, C.~Babin, R.~Nagy, I.~Gediz, E.~Hesselmeier, D.~Liu, M.~Joliffe,
  M.~Niethammer, D.~Dasari, V.~Vorobyov \emph{et~al.}, \enquote{Spin-controlled
  generation of indistinguishable and distinguishable photons from silicon
  vacancy centres in silicon carbide,} {\protect\JournalTitle{Nature
  communications}} \textbf{11}, 1--8 (2020).

\bibitem{Castelletto2019a}
S.~Castelletto, A.~S. {Al Atem}, F.~A. Inam, H.~J. von Bardeleben, S.~Hameau,
  A.~F. Almutairi, G.~Guillot, S.-i. Sato, A.~Boretti, and J.~M. Bluet,
  \enquote{{Deterministic placement of ultra-bright near-infrared color centers
  in arrays of silicon carbide micropillars},} {\protect\JournalTitle{Beilstein
  Journal of Nanotechnology}} \textbf{10}, 2383--2395 (2019).

\bibitem{Schaeper2022}
O.~Schaeper, Z.~Yang, M.~Kianinia, J.~E. Fr{\"{o}}ch, A.~Komar, Z.~Mu, W.~Gao,
  D.~N. Neshev, and I.~Aharonovich, \enquote{{Monolithic Silicon Carbide
  Metalenses},} {\protect\JournalTitle{ACS Photonics}} \textbf{9}, 1409--1414
  (2022).

\bibitem{Schuller2007}
J.~A. Schuller, R.~Zia, T.~Taubner, and M.~L. Brongersma, \enquote{Dielectric
  metamaterials based on electric and magnetic resonances of silicon carbide
  particles,} {\protect\JournalTitle{Phys. Rev. Lett.}} \textbf{99}, 107401
  (2007).

\bibitem{caldwell2013low}
J.~D. Caldwell, O.~J. Glembocki, Y.~Francescato, N.~Sharac, V.~Giannini, F.~J.
  Bezares, J.~P. Long, J.~C. Owrutsky, I.~Vurgaftman, J.~G. Tischler
  \emph{et~al.}, \enquote{Low-loss, extreme subdiffraction photon confinement
  via silicon carbide localized surface phonon polariton resonators,}
  {\protect\JournalTitle{Nano Lett.}} \textbf{13}, 3690--3697 (2013).

\bibitem{qiu2021quo}
C.-W. Qiu, T.~Zhang, G.~Hu, and Y.~Kivshar, \enquote{Quo vadis, metasurfaces?}
  {\protect\JournalTitle{Nano Letters}} \textbf{21}, 5461--5474 (2021).

\bibitem{solntsev2021metasurfaces}
A.~S. Solntsev, G.~S. Agarwal, and Y.~S. Kivshar, \enquote{Metasurfaces for
  quantum photonics,} {\protect\JournalTitle{Nature Photonics}} \textbf{15},
  327--336 (2021).

\bibitem{LinHassanfirooziJiangLiaoLeeWu+2022+2701+2709}
Y.-T. Lin, A.~Hassanfiroozi, W.-R. Jiang, M.-Y. Liao, W.-J. Lee, and P.~C. Wu,
  \enquote{Photoluminescence enhancement with all-dielectric coherent
  metasurfaces,} {\protect\JournalTitle{Nanophotonics}} \textbf{11}, 2701--2709
  (2022).

\bibitem{Alaee2018}
R.~Alaee, C.~Rockstuhl, and I.~Fernandez-Corbaton, \enquote{{An electromagnetic
  multipole expansion beyond the long-wavelength approximation},}
  {\protect\JournalTitle{Optics Communications}} \textbf{407}, 17--21 (2018).

\bibitem{Liu2019}
Z.~Liu, Y.~Xu, Y.~Lin, J.~Xiang, T.~Feng, Q.~Cao, J.~Li, S.~Lan, and J.~Liu,
  \enquote{{High- Q Quasibound States in the Continuum for Nonlinear
  Metasurfaces},} {\protect\JournalTitle{Physical Review Letters}} \textbf{123}
  (2019).

\bibitem{Shamkhi2019a}
H.~K. Shamkhi, K.~V. Baryshnikova, A.~Sayanskiy, P.~Kapitanova, P.~D. Terekhov,
  P.~Belov, A.~Karabchevsky, A.~B. Evlyukhin, Y.~Kivshar, and A.~S. Shalin,
  \enquote{{Transverse scattering and generalized kerker effects in
  all-dielectric mie-resonant metaoptics},} {\protect\JournalTitle{Physical
  Review Letters}} \textbf{122}, 193905 (2019).

\bibitem{Khokhar2022}
M.~Khokhar, F.~A. Inam, and R.~V. Nair, \enquote{{Kerker Condition for
  Enhancing Emission Rate and Directivity of Single Emitter Coupled to
  Dielectric Metasurfaces},} {\protect\JournalTitle{Advanced Optical
  Materials}} \textbf{10}, 2200978 (2022).

\bibitem{hulst1981light}
H.~C. Hulst and H.~C. van~de Hulst, \emph{Light scattering by small particles}
  (Courier Corporation, 1981).

\bibitem{mie1976contributions}
G.~Mie, \enquote{Contributions to the optics of turbid media, particularly of
  colloidal metal solutions,} {\protect\JournalTitle{Contributions to the
  optics of turbid media}} \textbf{25}, 377--445 (1976).

\bibitem{bezares2013mie}
F.~J. Bezares, J.~P. Long, O.~J. Glembocki, J.~Guo, R.~W. Rendell, R.~Kasica,
  L.~Shirey, J.~C. Owrutsky, and J.~D. Caldwell, \enquote{Mie
  resonance-enhanced light absorption in periodic silicon nanopillar arrays,}
  {\protect\JournalTitle{Optics Express}} \textbf{21}, 27587--27601 (2013).

\bibitem{bohren2008absorption}
C.~F. Bohren and D.~R. Huffman, \emph{Absorption and scattering of light by
  small particles} (John Wiley \& Sons, 2008).

\bibitem{evlyukhin2012demonstration}
A.~B. Evlyukhin, S.~M. Novikov, U.~Zywietz, R.~L. Eriksen, C.~Reinhardt, S.~I.
  Bozhevolnyi, and B.~N. Chichkov, \enquote{Demonstration of magnetic dipole
  resonances of dielectric nanospheres in the visible region,}
  {\protect\JournalTitle{Nano letters}} \textbf{12}, 3749--3755 (2012).

\bibitem{shilkin2017optical}
D.~A. Shilkin, M.~R. Shcherbakov, E.~V. Lyubin, K.~G. Katamadze, O.~S.
  Kudryavtsev, V.~S. Sedov, I.~I. Vlasov, and A.~A. Fedyanin, \enquote{Optical
  magnetism and fundamental modes of nanodiamonds,} {\protect\JournalTitle{ACS
  Photonics}} \textbf{4}, 1153--1158 (2017).

\bibitem{hinamoto2021menp}
T.~Hinamoto and M.~Fujii, \enquote{Menp: an open-source matlab implementation
  of multipole expansion for nanophotonics,} {\protect\JournalTitle{Osa
  Continuum}} \textbf{4}, 1640--1648 (2021).

\bibitem{Alaee2015}
R.~Alaee, R.~Filter, D.~Lehr, F.~Lederer, and C.~Rockstuhl, \enquote{{A
  generalized Kerker condition for highly directive nanoantennas},}
  {\protect\JournalTitle{Optics Letters}} \textbf{40}, 2645 (2015).

\bibitem{inam2011modification}
F.~Inam, T.~Gaebel, C.~Bradac, L.~Stewart, M.~Withford, J.~Dawes, J.~Rabeau,
  and M.~Steel, \enquote{Modification of spontaneous emission from nanodiamond
  colour centres on a structured surface,} {\protect\JournalTitle{New Journal
  of Physics}} \textbf{13}, 073012 (2011).

\bibitem{shamkhi2019transverse}
H.~K. Shamkhi, K.~V. Baryshnikova, A.~Sayanskiy, P.~Kapitanova, P.~D. Terekhov,
  P.~Belov, A.~Karabchevsky, A.~B. Evlyukhin, Y.~Kivshar, and A.~S. Shalin,
  \enquote{Transverse scattering and generalized kerker effects in
  all-dielectric mie-resonant metaoptics,} {\protect\JournalTitle{Physical
  review letters}} \textbf{122}, 193905 (2019).

\bibitem{Feifei2022PRL}
F.~Qin, Z.~Zhang, K.~Zheng, Y.~Xu, S.~Fu, Y.~Wang, and Y.~Qin,
  \enquote{Transverse kerker effect for dipole sources,}
  {\protect\JournalTitle{Phys. Rev. Lett.}} \textbf{128}, 193901 (2022).

\bibitem{Fuchs2015}
F.~Fuchs, B.~Stender, M.~Trupke, D.~Simin, J.~Pflaum, V.~Dyakonov, and G.~V.
  Astakhov, \enquote{{Engineering near-infrared single-photon emitters with
  optically active spins in ultrapure silicon carbide},}
  {\protect\JournalTitle{Nature Communications}} \textbf{6}, 7578 (2015).

\bibitem{Widmann2015}
M.~Widmann, S.~Y. Lee, T.~Rendler, N.~T. Son, H.~Fedder, S.~Paik, L.~P. Yang,
  N.~Zhao, S.~Yang, I.~Booker, A.~Denisenko, M.~Jamali, S.~{Ali Momenzadeh},
  I.~Gerhardt, T.~Ohshima, A.~Gali, E.~Janz{\'{e}}n, and J.~Wrachtrup,
  \enquote{{Coherent control of single spins in silicon carbide at room
  temperature},} {\protect\JournalTitle{Nature Materials}} \textbf{14},
  164--168 (2015).

\bibitem{InamCastnano22}
F.~A. Inam and S.~Castelletto, \enquote{Metal-dielectric nanopillar
  antenna-resonators for efficient collected photon rate from silicon carbide
  color centers,} {\protect\JournalTitle{Nanomaterials}} \textbf{13} (2023).

\bibitem{Mello2022}
O.~Mello, Y.~Li, S.~A. Camayd-Mu{\~{n}}oz, C.~DeVault, M.~Lobet, H.~Tang,
  M.~Lon{\c{c}}ar, and E.~Mazur, \enquote{{Extended many-body superradiance in
  diamond epsilon near-zero metamaterials},} {\protect\JournalTitle{Applied
  Physics Letters}} \textbf{120}, 061105 (2022).

\bibitem{singh1971nonlinear}
S.~Singh, J.~Potopowicz, L.~Van~Uitert, and S.~Wemple, \enquote{Nonlinear
  optical properties of hexagonal silicon carbide,}
  {\protect\JournalTitle{Applied Physics Letters}} \textbf{19}, 53--56 (1971).

\bibitem{frezza2017tutorial}
F.~Frezza, F.~Mangini, and N.~Tedeschi, \enquote{Tutorial: Introduction to
  electromagnetic scattering,} {\protect\JournalTitle{Journal of the Optical
  Society of America A}} \textbf{31}, 1--11 (2017).

\bibitem{Xu1999}
Y.~Xu, J.~S. Vu{\v{c}}kovi{\'{c}}, R.~K. Lee, O.~J. Painter, A.~Scherer, and
  A.~Yariv, \enquote{{Finite-difference time-domain calculation of spontaneous
  emission lifetime in a microcavity},} {\protect\JournalTitle{Journal of the
  Optical Society of America B}} \textbf{16}, 465 (1999).

\bibitem{Shaffer1971}
P.~T.~B. Shaffer, \enquote{{Refractive Index, Dispersion, and Birefringence of
  Silicon Carbide Polytypes},} {\protect\JournalTitle{Applied Optics, Vol. 10,
  Issue 5, pp. 1034-1036}} \textbf{10}, 1034--1036 (1971).

\end{thebibliography}
%\bibliography{Ashar}

%\end{thebibliography}

\end{document}